\documentclass[twocolumn,secnumarabic,amssymb, nobibnotes, aps, prd]{revtex4-1}

\usepackage{hyperref}
\usepackage{braket}
\usepackage{amsmath,amsfonts,amssymb,amsthm}
\usepackage{graphicx}
\usepackage{dcolumn}
\usepackage{bm}

\usepackage{lipsum}

\usepackage{algorithm,algcompatible,amsmath}
\usepackage{algpseudocode}

\begin{document}

\def\thesection{\Roman{section}}
\def\thesubsection{\Alph{subsection}}
\def\thesubsubsection{\arabic{subsubsection}}

\title{Supervised Learning Using a Dressed Quantum Network with \\ ``Super Compressed Encoding'': Algorithm and Quantum-Hardware-Based Implementation}%

\author{Saurabh Kumar}
\thanks{equal contribution}
 \affiliation{Department of Electrical Engineering, Indian Institute of Technology Delhi, New Delhi, Delhi 110016, India}
\author{Siddharth Dangwal}%
\thanks{equal contribution}
 \affiliation{Department of Electrical Engineering, Indian Institute of Technology Delhi, New Delhi, Delhi 110016, India} 
\author{Debanjan Bhowmik}
\email{debanjan@ee.iitd.ac.in}
\homepage{http://web.iitd.ac.in/~debanjan/}
\affiliation{Department of Electrical Engineering, Indian Institute of Technology Delhi, New Delhi, Delhi 110016, India}%

\begin{abstract}
Implementation of variational Quantum Machine Learning (QML) algorithms on Noisy Intermediate-Scale Quantum (NISQ) devices is known to have issues related to the high number of qubits needed and the noise associated with multi-qubit gates. In this paper, we propose a variational QML algorithm using a dressed quantum network to address these issues.  Using the ``super compressed encoding'' scheme that we follow here, the classical encoding layer in our dressed network drastically scales down the input-dimension, before feeding the input to the variational quantum circuit. Hence, the number of qubits needed in our quantum circuit goes down drastically. Also, unlike in most other existing QML algorithms, our quantum circuit consists only of single-qubit gates, making it robust against noise. These factors make our algorithm suitable for implementation on NISQ hardware. \\
\hspace*{1em}To support our argument, we implement our algorithm on real NISQ hardware and thereby show accurate classification using popular machine learning data-sets like Fisher's Iris, Wisconsin's Breast Cancer (WBC), and Abalone. Then, to provide an intuitive explanation for our algorithm's working, we demonstrate the clustering of quantum states, which correspond to the input-samples of different output-classes, on the Bloch sphere (using WBC and MNIST data-sets). This clustering happens as a result of the training process followed in our algorithm. Through this Bloch-sphere-based representation, we also show the distinct roles played (in training) by the adjustable parameters of the classical encoding layer and the adjustable parameters of the variational quantum circuit. These parameters are adjusted iteratively during training through loss-minimization.
\end{abstract}


\maketitle

\section{Introduction}


\vspace{0.4em}
Quantum computing \cite{Preskill2018, vijay2, google2018} has been considered attractive of late for implementing Machine Learning (ML) algorithms for several applications related to Artificial Intelligence \cite{lecun2015,biamonte2017_review, woss2018, havlivcek2019}. Variational algorithms form a special class of such Quantum Machine Learning (QML) algorithms \cite{havlivcek2019, schuld2019, benedetti2019, CQC2019, adhikary2020, salinas2020, mari2019}. Noisy Intermediate-Scale Quantum (NISQ) devices are considered suitable candidates for implementing such algorithms \cite{Preskill2018, ibmq, Grollier_QNeuro2020, havlivcek2019}.

\vspace{0.4em}
However, there are various challenges associated with the implementation of most variational QML algorithms on NISQ hardware:

\begin{enumerate}

\item
Due to physical constraints, NISQ hardware mostly allows either single-qubit or two-qubit gates  \cite{google2018, ibmq}. But most variational QML algorithms need multi-qubit gates in the variational/ parametrized quantum circuits that they use \cite{havlivcek2019, benedetti2019}. So any gate-operation involving more than two qubits needs to be decomposed into a series of single-qubit and two-qubit gates. Such design increases the gate-count as well as the circuit-depth while implementing such QML algorithms in hardware. An increase in the gate-count leads to an increased accumulation of noise in the form of gate-errors \cite{NASAreview}. A larger circuit-depth implies a longer execution time and, hence, higher decoherence \cite{ibmq}. The gate-count and the circuit-depth further increase due to the constrained architecture of the NISQ devices. Limited connectivity between physical qubits implies that not all two-qubit gates are implementable directly. To realise such ``forbidden'' gates, one needs to remap logical qubits to physical qubits through SWAP insertion \cite{saber2019,codar2020,swap2020,algo_opt_2019}.

\item
The existing QML algorithms mostly follow the amplitude encoding scheme or the qubit encoding scheme to encode the features of each input-sample (in the ML data-sets) as qubits for the variational/ parametrized quantum circuit \cite{havlivcek2019,schuld2019,Grollier_QNeuro2020,Schwab_QubitEncoding}. As a consequence of these schemes, the number of qubits in the quantum circuit depends on the dimension of the input-samples in a given data-set; this dimension is typically high \cite{havlivcek2019,schuld2019,Grollier_QNeuro2020,Schwab_QubitEncoding}. This makes the physical implementation of such QML algorithms difficult since numerous qubits are needed \cite{NASAreview}.

\end{enumerate}

  \begin{figure*}[!t]
    
      \includegraphics[width=0.9\textwidth]{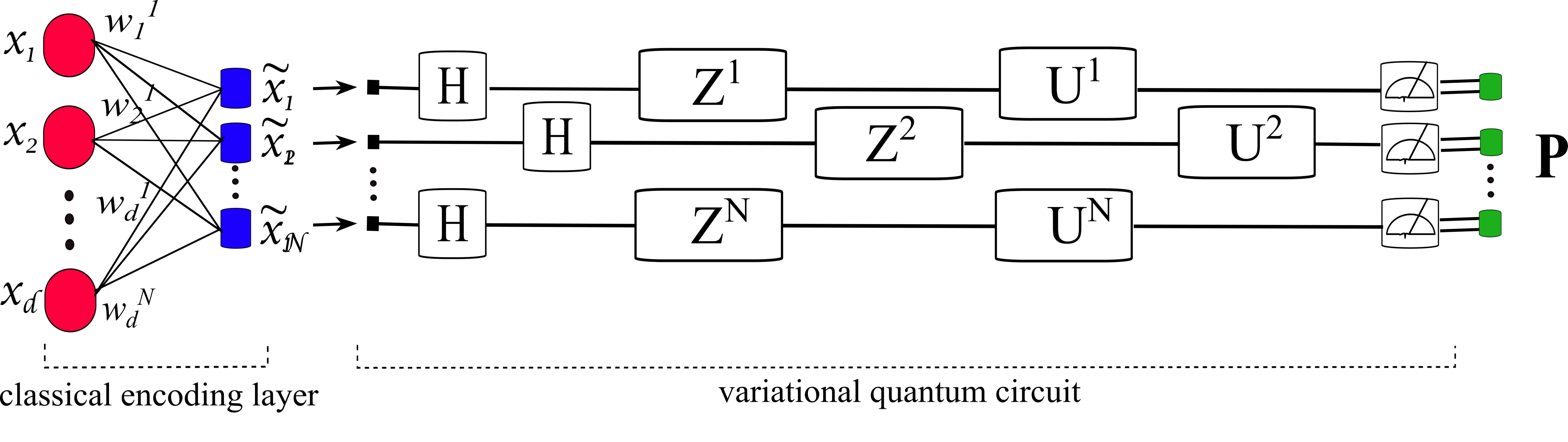}
      \caption{The structure of the dressed quantum network we design here is shown in this schematic. The classical encoding layer converts the input $(x_{1},x_{2},...x_d)$ to {($\Tilde{x}_1$,$\Tilde{x}_2$,...$\Tilde{x}_N$)}, where $d$ is the dimension of the input-sample and $N$ is the number of possible output classes/ labels of the input-sample. Parameters of the classical encoding layer ($w_i^j$; $i = 1, 2, \cdots, d$ ; $j = 1, 2, \cdots, N$) are updated after every epoch during the training of the dressed quantum network, following our proposed algorithm. In the variational quantum circuit, the superscripts in the labels of the gates indicate the qubit on which it operates. $Z^j = e^{i \sigma_3 \Tilde{x}_j}$, $U^j = (e^{i \sigma_3 \alpha_1^j} e^{i \sigma_2 \alpha_2^j} e^{i \sigma_3 \alpha_3^j})$. $U^j$ corresponds to the SU(2) operations on the $j$-th qubit. $\alpha_1^j$,$\alpha_2^j$, and $\alpha_3^j$ are the SU(2) parameters/ rotation parameters for the $j$-th qubit. For every qubit from $j$=1 to $j=N$, the SU(2) parameters are also updated after every epoch during the training of the dressed quantum network. Projective measurement is carried out on each qubit for the outcome $\sigma_3=+1$. The set of probabilities for all the qubits is given by ${\bf P}= (P^{1},P^{2},...P^{N})$.}
      \label{fig:scematic} 
     \end{figure*}




\vspace{0.4em}
In this paper, we propose a variational QML algorithm using a dressed quantum network \cite{adhikary2020, salinas2020, mari2019} (Fig. ~\ref{fig:scematic}). We use an aggressive encoding scheme in our network, which we call ``super compressed encoding.'' We use this scheme instead of the amplitude encoding scheme or the qubit encoding scheme mentioned earlier \cite{havlivcek2019,schuld2019,Grollier_QNeuro2020,Schwab_QubitEncoding}. Following our scheme, the classical encoding layer in our dressed network scales down the input-dimensions drastically. Independent of the number of features/ dimensions that each input-sample in an ML data-set has, the input-sample is encoded into a one-dimensional scalar for every possible output-class (see the details in Section II). The parameters used for this encoding are adjustable (Fig. ~\ref{fig:scematic}).

\vspace{0.4em}
For every output-class, corresponding to that one-dimensional scalar, a qubit-state is prepared. Several parametrized (these parameters are also adjustable) single-qubit gates are applied on that qubit in the variational quantum circuit (Fig. ~\ref{fig:scematic}). Then a measurement is performed to generate the loss based on the output-class that the input sample belongs to; we follow a supervised learning scheme here \cite{lecun2015}. Our algorithm is a classical-quantum hybrid variational algorithm \cite{mari2019,havlivcek2019,adhikary2020}. So all the adjustable parameters in our network are updated classically and iteratively over several epochs such that the loss decreases after every epoch (until the loss doesn't decrease any further/ we achieve convergence). Thus the network gets trained (find the details in Section II and in the block titled `Algorithm').

\vspace{0.4em}
As a result of using this ``super compressed encoding'' scheme, our proposed algorithm addresses the issues related to the existing variational QML algorithms mentioned above. The number of qubits does not depend on the input-dimensions. It only depends on the number of output-classes, which is typically much lower than the number of input dimensions. Hence, we need fewer qubits for our implementation (find a quantitative comparison in Section V). Also, since we eliminate the need for multi-qubit gates, our algorithm is much more robust against noise \cite{ibmq}.

\vspace{0.4em}
In Section III of the paper, we implement our proposed algorithm on three separate platforms: a classical computer where the steps in our algorithm are carried out through Python-based programming, a quantum computation software called Qiskit, and real NISQ hardware (IBM-Q). On each of these platforms, we show that our classifier, despite using our aggressive encoding scheme (`super compressed encoding'), can indeed classify samples from popular ML data-sets such as Fisher's Iris, Wisconsin Breast Cancer (WBC) (diagnosis), and Abalone with high accuracy (Fig. ~\ref{fig:loss_min}, Table ~\ref{tab:accuracy}). From our IBM-Q-based implementation, we also show that our algorithm is robust against noise (Fig. ~\ref{ibmq_graph}, Table ~\ref{tab:ct}).

\vspace{0.4em}
In Section IV, we intuitively explain how the classification occurs in our network by representing the qubit-state, which we use, on the Bloch sphere as it evolves during the implementation of our proposed algorithm. Following this method, we show the clustering of qubit-states, which correspond to the input-samples belonging to different output-classes, on the Bloch sphere. This clustering happens as a result of the training process followed in our algorithm. Using the Bloch-sphere-based representation for binary classification (2 output-classes) on the WBC data-set and the MNIST data-set of handwritten digits (Table ~\ref{table_mnist}), we explain the training process intuitively. We also show the distinct roles played, during the training process, by the adjustable parameters in the classical encoding layer and that in the quantum circuit of our designed dressed quantum network (Fig. ~\ref{fig:visual}, Fig. ~\ref{fig:visual_MNIST}). 

\vspace{0.4em}
To the best of our knowledge, such a Bloch-sphere-based approach has not been used before to show the evolution of quantum states (corresponding to the input-samples) as they are acted upon by the quantum gates and thus explain the working of other existing QML algorithms. But extensive research has been carried out recently to explain the internal mechanism behind classical ML algorithms' working \cite{ExplainableAI1,ExplainableAI2}. So, in that context, our Bloch-sphere-based explanation of the working of our QML algorithm may be considered very relevant for research on QML algorithms in general.

\vspace{0.4em}
In Section V, we compare the data-sets used by us with that used by other existing QML algorithms (Table ~\ref{table_datasetcomp}). We show that our algorithm can handle ML data-sets as complex or more complex than those handled so far by the different existing QML algorithms. Then we explore, in more detail, the advantages of our algorithm (robustness against noise, low number of qubits, etc., as mentioned above) compared to other existing QML algorithms through quantitative estimates. In Section VI, we conclude the paper.

\vspace{0.4em}
Though the same encoding scheme as the one used here has been used in \cite{adhikary2020}, the QML algorithm there uses a multi-level quantum system (qu-N-it) for multi-class classification. But for implementation on a practical NISQ hardware like IBM-Q, only a 2-level quantum system or qubit can be used. Hence, the algorithm in \cite{adhikary2020} can only be used for binary classification. But in this paper, we extend such qubit-based algorithm to the case of an arbitrary number of possible output-classes (Fig. ~\ref{fig:loss_min}, Table ~\ref{tab:accuracy}). Also, here we show the implementation of our algorithm on Qiskit and IBM-Q, unlike in \cite{adhikary2020}. Moreover, unlike in \cite{adhikary2020}, here we explain the working of our algorithm through Bloch-sphere-based representation.

\section{The Proposed Algorithm}

\subsection{Classical Encoding Layer}

\vspace{0.4em}
The dressed quantum network that we design corresponding to our proposed algorithm is shown in Fig. ~\ref{fig:scematic}. In our network, much like in \cite{mari2019}, the classical encoding layer, acting on the input, is used to prepare the qubits for the variational quantum circuit.

\vspace{0.4em}
Let us consider a data-set $\mathcal{S} = \{ ({\bf x}, f({\bf x}))\}$. Each entry in $\mathcal{S}$ is an ordered pair. It consists of a sample, represented as a vector ${\bf x}= (x_{1},x_{2},...x_d) \in \mathbb{R}^d$ and its associated label $f({\bf x})$. The label corresponds to the output-class that the sample belongs to. Thus, $f$ maps each sample to one of the $N$ labels in the set: $\mathcal{L} = \{ l_1, l_2, \cdots, l_N\}$; $f: {\bf x} \rightarrow \mathcal{L}$. The mappings that we consider here are many-to-one.

\vspace{0.4em}
In our ``super compressed encoding'' scheme, our classical encoding layer has the input layer with $d$ nodes connected with an output layer of $N$ nodes; there is no hidden layer (Fig.~\ref{fig:scematic}) \cite{lecun2015,adhikary2020} . $d$ is the dimension of each input-vector/ input-sample; $N$ represents the total number of possible output-classes. We denote this classical transformation as ${\cal N}_{d \rightarrow N}$. It corresponds to a simple Vector Matrix Multiplication (VMM) operation ${\cal N}: {\bf x} \rightarrow {\bf \Tilde{x}}$; $\Tilde{x}_j = \sum_i x_i w_i^j$; $i = 1, 2, \cdots, d$ ; $j = 1, 2, \cdots, N$. Thus,  $w_i^j$ represents an element of a $N \times d$ dimensional weight matrix $W$. The transformed vector ${\bf \Tilde{x}}$ is $N$-dimensional. $N$ is typically much smaller than the dimension of the original input vector ($N < < d$). 

\vspace{0.4em}
Thus, we observe here that following our ``super compressed encoding'' scheme, each input-sample, independent of its original dimensions ($d$), is drastically reduced to a one-dimensional scalar for each of the $N$ output-classes. In Section I, we already highlight this point.


\subsection{Variational Quantum Circuit}

\vspace{0.4em}
In the variational quantum circuit connected to the classical encoding layer (Fig. ~\ref{fig:scematic}), the compressed data $\Tilde{x}_j$ ($j = 1, 2, \cdots, N$) is encoded into a $N$-qubit quantum state $\vert \psi({\bf x})\big>$ through the following gate operations:
\begin{equation}
\label{eq:state_prep}
\ket{ \psi(\bf x)} = \otimes_{j = 1}^{N} e^{i \sigma_3 \Tilde{x}_j} H^j  \ket{0}.
\end{equation}
Here, $H$ and $\sigma_3$ are the Hadamard gate and the third Pauli matrix, respectively, defined in a two-dimensional Hilbert space. The index $j$ labels individual qubits.

\vspace{0.4em}
Following the state-preparation step, we add a layer of parametrized SU(2) operations on each qubit \cite{adhikary2020,Debanjan2009}. In \cite{adhikary2020}, if a 2-level quantum system is used, a similar SU(2) operation is applied. But then, the overall network in \cite{adhikary2020} has only one qubit. In that case, the algorithm in \cite{adhikary2020} can only be used only for the particular case of binary classification. But in this paper, the number of output-classes can be greater than 2 ($N \geq 2$), as shown in Section III.

\vspace{0.4em}
Here, in this paper, the SU(2) operations (Fig. ~\ref{fig:scematic}) lead to the transformation:
\begin{equation}
\label{eq:su2trans}
\ket{\bar \psi(\bf x)} = (\otimes_{j = 1}^{N} e^{i \sigma_3 \alpha_1^{j}} e^{i \sigma_2 \alpha_2^{j}} e^{i \sigma_3 \alpha_3^{j}}) \ket{\psi(\bf x)}
\end{equation}
where $\alpha_1^{j}$, $\alpha_2^{j}$ and $\alpha_3^{j}$ are the rotation parameters of the SU(2) operation on the $j$-th qubit. It is to be noted that all the gates used in equation ~\ref{eq:state_prep}
 and equation ~\ref{eq:su2trans} are single-qubit gates, as mentioned in Section I (Fig.~\ref{fig:scematic}). 
 
\vspace{0.4em}
The final layer added to the circuit performs a projective measurement (Fig. ~\ref{fig:scematic}). This step is the read-out step; we use the information obtained from this step directly to classify the data. The measurement of our choice is the projection operator for the outcome: $\sigma_3 = +1$. We record the probability for the outcome $\sigma_3 = +1$ for each qubit and store them as a $N$-dimensional probability vector ${\bf P}= (P^{1},P^{2},...P^{N})$. It is to be noted that we do not have a classical layer  (hence no adjustable parameters) on the network's output-side, unlike the dressed quantum network in \cite{mari2019}.

\vspace{0.4em}
We subsequently use the probability vector ${\bf P}$ to compute the loss function, as we discuss next \cite{bishop1995}.

\subsection{The Loss-Computation and the Training Process}


\vspace{0.4em}
Since we follow the supervised learning scheme \cite{lecun2015}, we already know the class that each sample (\textbf{x}) belongs to ($f: {\bf x} \rightarrow \mathcal{L}$) and hence the target probability vector for that sample. We set the target probability vector for any sample of the $k$-th class (${\bf P}_{target}^{(k)}$) as:
  \begin{equation}
  \label{target}
      P_{target}^{(k) s} =
    \begin{cases}
      1 & \text{if $s = k$}\\
      0 & \text{otherwise}
    \end{cases}
  \end{equation}

\vspace{0.4em}
Here, $ P_{target}^{(k) s}$ is the $s$-th element of the vector ${\bf P}_{target}^{(k)}$. To be classified correctly, the probability vector  ${\bf P}$ (consequence of the measurement as described above) for every input belonging to the $k$-th class must evolve to ${\bf P}_{target}^{(k)}$, given by equation ~\ref{target}. Minimization of a properly constructed loss function leads to that.

\vspace{0.4em}
Comparing the two probability vectors above, the cross-entropy loss for a training sample belonging to the $k$-th class follows straightaway \cite{bishop1995}:

     
\begin{equation}
\label{crossentropy}
{\cal E}_{CrossEntropy} = - \sum_{s = 1}^{N} P_{target}^{(k)s} \log_{e} \Sigma^s({\bf P}).
\end{equation} 
$\Sigma(\cdot)$ in equation~\ref{crossentropy} is the SoftMax function;  $\Sigma^s({\bf P}) = e^{P^s} / (\sum_{s'} e^{P^{s'}}) $. 

\vspace{0.4em}
In every epoch, the loss for each training sample is calculated once. The total loss for an epoch is the sum of the individual losses for each training sample calculated in that epoch.


\vspace{0.4em}
Our dressed quantum network is trained for data-classification by minimizing the loss function, after every epoch, with respect to the weights of the classical encoding layer ($w_i^j$; $i = 1, 2, \cdots, d$ ; $j = 1, 2, \cdots, N$) and the parameters in the SU(2) operation ($\alpha_1^{j}$, $\alpha_2^{j}$ and $\alpha_3^{j}$; $j = 1, 2, \cdots, N$). These adjustable parameters are initialized to random values at the beginning of the training. Then during the training process, they are adjusted once every epoch (as mentioned in Section I) such that the loss decreases after every epoch (until the loss doesn't decrease any further/ we achieve convergence) and the probability vector for a sample belonging to the $k$-th class evolves, over many epochs, into the target probability vector for that class. Thus all these parameters are adjusted iteratively to train our dressed network.

\vspace{0.4em}
It is to be noted here that our algorithm is also different from \cite{mari2019} in the following way. In \cite{mari2019}, most parameters of the classical network, connected to the input, are pre-tuned and fixed. They do not change when the parameters of the quantum circuit are updated. But in our algorithm, we iteratively adjust all the parameters of our classical encoding layer during the training process along with adjusting the SU(2) parameters in our variational quantum circuit. Also, the number of parameters used in the classical network of \cite{mari2019} is much higher than the number of parameters we use in our classical encoding layer, here.

\vspace{0.4em} 
\textit{We summarize our proposed algorithm for training our dressed quantum network in the block titled ``Algorithm'' for easy reference.}

\vspace{0.4em} 
To determine the classification accuracy, we use a simple classification metric. A sample ${\bf x}$ belonging to the $k$-th class is said to  be correctly classified if the $k$-th element of the probability vector ${\bf P}$ is greater than all other elements in that vector. To sharpen the criterion, we further impose the condition that the value of the $k$-th element in ${\bf P}$ must exceed a certain threshold $c_t$. Greater the value of $c_t$, more stringent is the classification criterion. Thus, $c_t$ is the classification-metric in our algorithm.

\vspace{0.4em} 
In the next section (Section III), we implement our proposed algorithm on different platforms and show our classification results, using this algorithm, on different data-sets. The distinct roles played, during the training process, by the adjustable weights of the classical encoding layer ($w_i^j$; $i = 1, 2, \cdots, d$ ; $j = 1, 2, \cdots, N$) and the adjustable SU(2) parameters ($\alpha_1^{j}$, $\alpha_2^{j}$ and $\alpha_3^{j}$; $j = 1, 2, \cdots, N$) in the variational quantum circuit are explained in Section IV.

\begin{algorithm}[H]
   \caption{Our Proposed QML Algorithm}
    \label{Algorithm}
      \begin{algorithmic}[1]
          \State \textbf{Input encoding (data compression):} For each input vector/ sample ${\bf x}= (x_{1},x_{2},...x_d)$, use the classical encoding layer (${\cal N}: {\bf x} \rightarrow {\bf \Tilde{x}}$), connected to the input, to generate $\Tilde{x}_j = \sum_i x_i w_i^j$; $i = 1, 2, \cdots, d$ ; $j = 1, 2, \cdots, N$. $N$ is the number of possible output classes/ labels of the input (Fig. ~\ref{fig:scematic}).
           \State \textbf{Input encoding (qubit preparation):} $\Tilde{x}_j$  ($j = 1,2, \cdots, N$) is encoded in $N$-qubit state: $|\psi(\mathbf{x})\rangle=$ $\otimes_{j=1}^{N} e^{i \sigma_{3} \tilde{x}_{j}} H^{j}|0\rangle^{j}$( equation ~\ref{eq:state_prep}, Fig. ~\ref{fig:scematic}).
      \State \textbf{SU(2) operation:} $ \ket{\bar \psi (\bf x)} = (\otimes_{j = 1}^{N} e^{i \sigma_3 \alpha_1^{j}} e^{i \sigma_2 \alpha_2^{j}} e^{i \sigma_3 \alpha_3^{j}}) \ket{ \psi (\bf x)}$ (equation ~\ref{eq:su2trans}, Fig. ~\ref{fig:scematic}).
1)
      \State \textbf{Measurement:} Projective measurement is carried out on each qubit for the outcome: $\sigma_3 = +1$. The set of probabilities for all the qubits in \(\ket{\bar \psi(\bf x)}\) is given by ${\bf P}= (P^{1},P^{2},...P^{N})$ (Fig. ~\ref{fig:scematic}).
      \State \textbf{Computation of loss:} Loss is calculated for each sample after comparing the probability vector ${\bf P}= (P^{1},P^{2},...P^{N})$, obtained from the measurement, with the target probability vector for the class that the sample belongs to (equation ~\ref{target}, ~\ref{crossentropy}).  
        \State \textbf{Optimization of Loss} Steps 1–5 are repeated for all input-samples in the training data-set. The loss for all the samples is added to generate the total loss for that epoch. Then the loss is minimized classically by updating the adjustable parameters: the weight parameters of the classical encoding layer  ($w_i^j$; $i = 1, 2, \cdots, d$ ; $j = 1, 2, \cdots, N$) and the the SU(2) parameters ($\alpha_1^{j}$, $\alpha_2^{j}$ and $\alpha_3^{j}$; $j = 1, 2, \cdots, N$).
     \State The above process is repeated over several epochs (loss for each sample in the training data-set is calculated once per epoch) until we converge to the minimum loss for all the samples in the training data-set.
   \end{algorithmic}
\end{algorithm}

 \section{Implementation of the Algorithm and Data Classification Using It}

\subsection{Implementation on Three Separate Platforms and Classification-Results}
 
\begin{figure*}[!t]
    \centering
    \includegraphics[width= 0.9
    \textwidth]{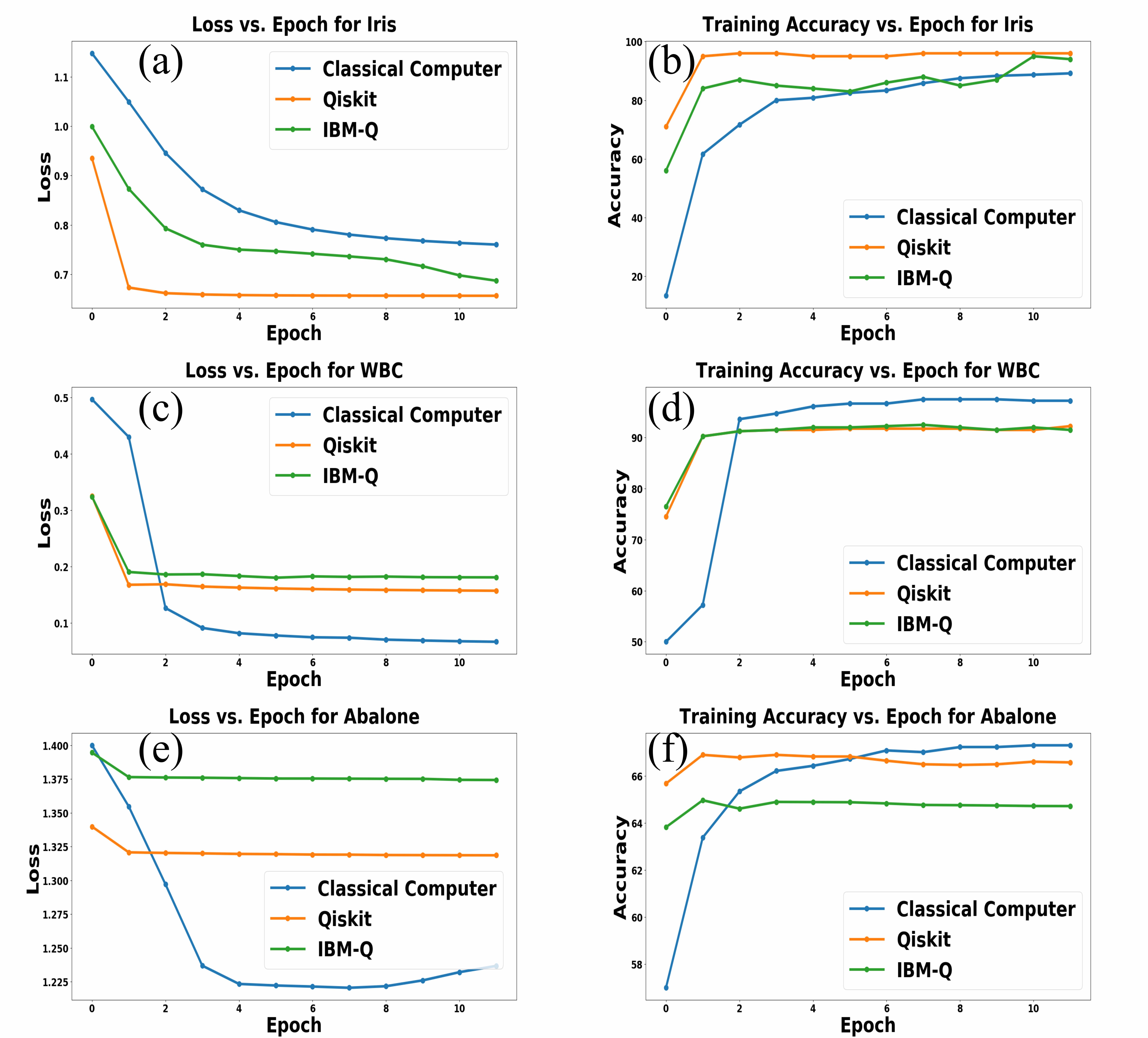}
    \caption{\label{fig:loss_min} Plots of the loss (Eq.~\ref{crossentropy}) as a function of epochs during training for (a) the Fisher's IRIS data-set, (c) the WBC data-set, and (e) the Abalone data-set. Accuracy on the training data-set (training accuracy) has been plotted as a function of epochs in (b) for the Fisher's Iris data-set, (d) for the WBC data-set, and (f) for the Abalone data-set. For this purpose, the algorithm has been implemented on three different platforms: 1. programming on a classical computer (Python code) 2. Qiskit \cite{Qiskit} 3. quantum hardware (IBM-Q) \cite{ibmq}. The value of the classification-metric $c_t$  ($c_t$ is defined in Section II) is chosen to be 0.5, for the accuracy-calculation.} 
\end{figure*} 

\begin{table*}[!t]
\begin{center}
\begin{tabular}{| l | l | l | l | l |}  
\hline
{\bf Dataset} & {\bf Classification type} &{\bf Classical computer} & {\bf Qiskit} & {\bf IBM-Q} \\ 

{} & {} & {\bf (Python code)} & {} & {} \\ \hline
Fisher's Iris & 3 classes & 90$\%$ &  94$\%$ &  82$\%$  \\ \hline
WBC & 2 classes &92.37$\%$ & 96.45$\%$ & 91.71$\%$  \\ \hline
Abalone & 6 classes & 67.70$\%$ & 67.44$\%$ & 67.22$\%$  \\ \hline
\end{tabular}
\end{center}
\caption{\label{tab:accuracy} Classification accuracy numbers on the test data-sets (test accuracy) as obtained by implementing our algorithm on three different platforms: 1. programming on a classical computer (Python code) 2. Qiskit \cite{Qiskit} 3. quantum hardware (IBM-Q) \cite{ibmq}. The value of the classification-metric $c_t$  ($c_t$ is defined in Section II) is chosen to be 0.5, for the accuracy-calculation.}
\end{table*}

\vspace{0.4em}
To assess our classifier, we have executed our algorithm for multi-class ($N \geq 2$) classification, discussed above, for three benchmark data-sets: 

\begin{enumerate}

\item Fisher's Iris data-set: 4-dimensional input ($d=4$), 3 output-classes/ labels ($N=3$) \cite{dua2019uci}

\item Wisconsin Breast Cancer (WBC) data-set:  30-dimensional input ($d=30$),  2 output-classes/ labels ($N=2$)  \cite{dua2019uci}

\item Abalone data-set: 8-dimensional input ($d=8$), 6 output-classes/ labels ($N=6$)     \cite{dua2019uci}.

\end{enumerate}

More details on these data-sets and how we have used them here can be found in Appendix A below.

\vspace{0.4em}
We have implemented our algorithm, using the above data-sets, on three separate platforms:

\begin{enumerate}
    \item A classical computer, using Python programming language: The analytic expressions in equation ~\ref{eq:state_prep}–\ref{crossentropy} are explicitly used in our codes.
    
    \item Qiskit, a quantum computing software: We have performed a noiseless simulation of the variational quantum circuit in our algorithm on this platform (Fig. ~\ref{fig:scematic}) \cite{Qiskit}. The classical encoding layer, the calculation of the loss, and the loss minimization are all implemented on a classical computer since our algorithm is a classical-quantum hybrid variational algorithm \cite{mari2019,havlivcek2019,adhikary2020}. However, for the classical optimization, a package provided by the Qiskit (Penny Lane) is used in this case \cite{pennylane2018}, unlike in platform 1, where we wrote our own Python code for the optimization.
    
    \item Quantum hardware (IBM-Q) \cite{ibmq}: We have used three separate systems to implement the variational quantum circuit in our algorithm: IBM-Q Rome, IBM-Q Armonk, and IBM-Q Melbourne. A different system is used for a different data-set (see Appendix B below for our reason behind such choices). For this kind of implementation, first, noisy simulation is performed on Qiskit. Noise-models, typical to IBM-Q Rome, IBM-Q Armonk, and IBM-Q Melbourne, are used.  After every training epoch, the model-parameters are recorded. Then, these parameter-values are fixed in the variational quantum circuit, both in the noisy Qiskit simulator as well as in the real IMB-Q hardware. The outcomes of the model (probabilities) from both the platforms are compared, for randomly sampled training data. An excellent agreement between the two platforms' results is observed, thus establishing an equivalence between the two platforms. More details about the implementation can be found in Appendix B below. Here also, the classical encoding layer, the calculation of the loss, and the loss minimization are all implemented on a classical computer.
    
\end{enumerate}

\vspace{0.4em}
The loss (as determined by equation ~\ref{crossentropy}) and the classification accuracy on the training data-set (training accuracy) are plotted as functions of epochs in Fig. \ref{fig:loss_min} for the different data-sets and the different platforms. The plots show convergence in training for all the cases. The value of the classification-metric $c_t$  ($c_t$ is defined in Section II) is chosen to be 0.5, for the accuracy-calculation.

\vspace{0.4em}
 Our classification accuracy results on the test data-set (test accuracy) are summarized in Table \ref{tab:accuracy} for classification-metric $c_t = 0.5$ ($c_t$ is defined in Section II). More information on how we split each data-set into a training data-set and a test data-set can be found in Appendix A below. The accuracy numbers obtained in Table \ref{tab:accuracy} are close to the accuracy numbers obtained from classical ML and deep learning algorithms for the same data-sets \cite{salama2012experimental,pinto2018iris,sahin2018abalone,khalifa2019single}. This shows that our ``super compressed encoding'' is effective despite aggressively reducing the input-dimensions.

\vspace{0.4em}
The classification accuracy (test) for the same data-set varies across the different platforms in Table \ref{tab:accuracy}. This can be attributed to the fact that the probability vector ${\bf P}$, based on which the classification occurs, is not identical when evaluated on different platforms. Two key factors are responsible for this behavior:

\begin{enumerate}
     \item The analytic expression of ${\bf P}$, which is explicitly used in platform 1, assumes that we have access to an infinitely large number of identically prepared states $\ket{\bar \psi (\bf x)}$. A projective measurement is performed on each of these states, and ${\bf P}$ is calculated from the resultant statistics. In both Qiskit and IBM-Q, the projective measurement is performed only for a finite number of times. This leads to a mismatch between the probability values obtained from platform 1 and those that are obtained from the other two platforms. This affects the loss function and hence the training. So it also affects the overall accuracy calculation. Interestingly, Qiskit (platform 2) offers the highest accuracy among the three platforms for most data-sets. This may be due to the high efficiency of the inbuilt optimizers (for the loss function) that the Penny Lane package (of Qiskit) uses. The Python code that we have written ourselves for optimization on platform 1 may have lower efficiency than that. Noise from actual quantum hardware affects the accuracy number for platform 3 adversely, as we discuss in the next point.
     
    
    \item The IBM-Q devices, used in our implementation (platform 3), suffer from noise in forms of gate-error, decoherence noise, etc. This results in imperfect preparation of $\ket{\bar \psi (\bf x)}$ and imperfect projective measurement. The effect of noise is not accounted for in the other two platforms; it leads to different ${\bf P}$ values for platform 1 and 2 compared to platform 3. The effect of noise, inherent in the real quantum hardware, can be seen in Table \ref{tab:accuracy}. The noise inevitably reduces the classification accuracy for the real quantum hardware compared to the other two platforms for a given data-set in most cases. Nonetheless, our accuracy numbers are still high for the real quantum hardware. Thus, we show that our algorithm is quite robust against noise. 
    
\end{enumerate}


\subsection{Robustness against noise}

 \begin{figure}
    \centering
    \includegraphics[width = 0.5 \textwidth]{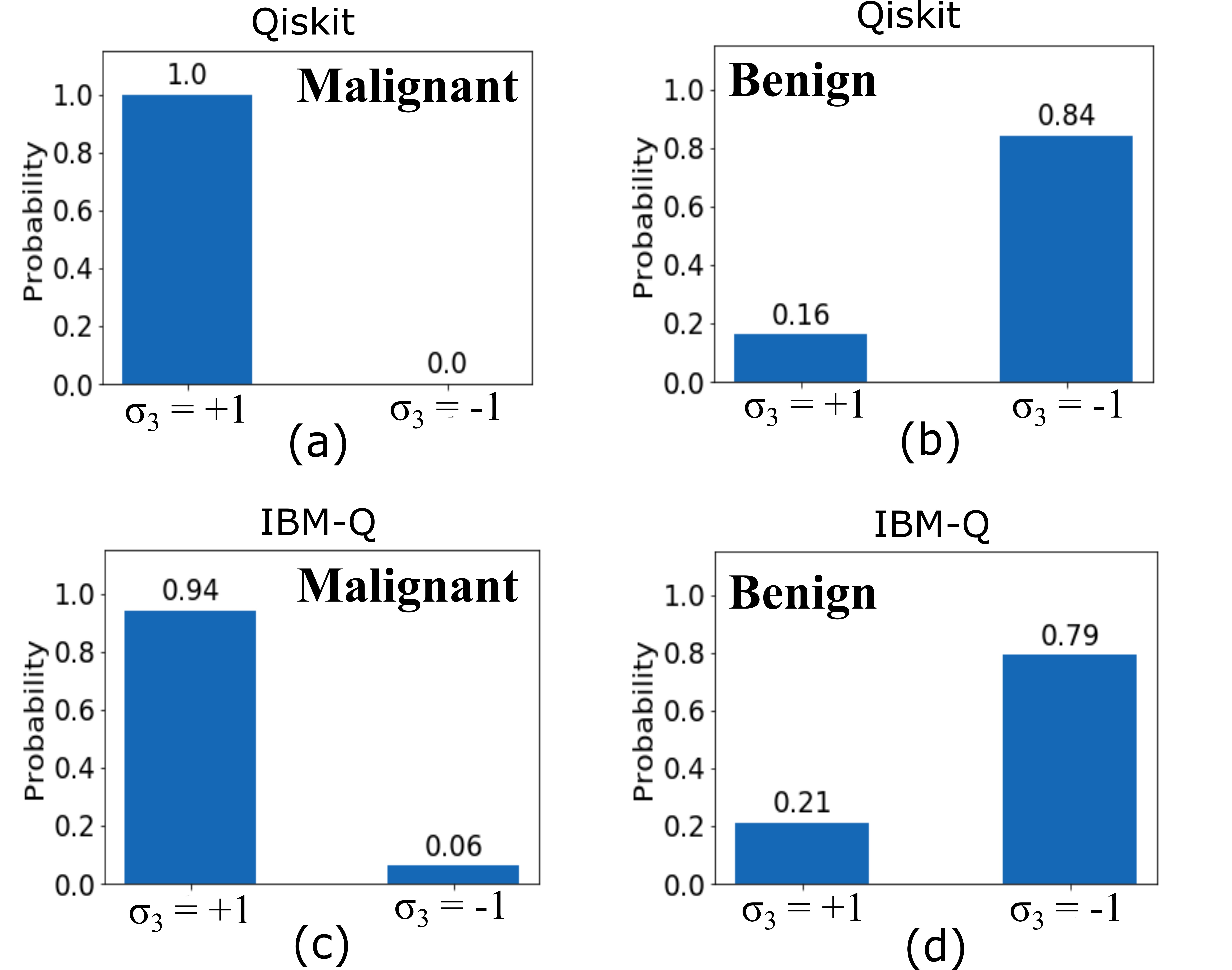}
    \caption{\label{ibmq_graph} The probabilities of the outcome of a projective measurement on a qubit being $\sigma_3 = +1$  ($P_+$) and the outcome being $\sigma_3 = -1$ ($1-P_+$) are obtained from a trained classifier (modified algorithm) for a sample labelled as ``malignant'' (a,c) and a sample labelled as ``benign'' (b,d) in the WBC dataset. The results in (a,b) are for the Qiskit-based implementation (platform 2) while those in (c,d) are for the IBM-Q-based implementation (platform 3). }
    
\end{figure}

\vspace{0.4em}
To further elaborate on the robustness of our implementation to noise, we show the results for the implementation of our algorithm on Qiskit (platform 2) and IBM-Q hardware (platform 3) for the WBC data-set in more detail (Fig.~\ref{ibmq_graph}). Since it is a 2-class-dataset, two qubits need to be used following our algorithm. But the information from one qubit is enough for classification if the network is properly trained. This is because after training, when the sample belongs to the ``malignant'' category, the first qubit is expected to evolve to $\ket{0}$, and the second qubit is expected to evolve to $\ket{1}$ (equation ~\ref{target}). Similarly, when the sample belongs to the ``benign'' category, the first qubit is expected to evolve to $\ket{1}$, and the second qubit is expected to evolve to $\ket{0}$. Thus after correct training, as we can see from here, the second qubit offers redundant information.

\vspace{0.4em}
So the loss function in equation ~\ref{crossentropy} can be reformulated in this specific case of binary classification to account for projective measurement on only one of the two qubits. Let the probability that the outcome of a projective measurement ($\sigma_3 = +1$) on the first qubit be denoted by $P_+$; probability of the outcome being ($\sigma_3 = -1$) on the same qubit $= 1-P_+$. In that case, loss function turns out to be: 

\begin{equation}
 \label{error_lin_bin}
     {\cal E} = \sum_{p = 1}^{m_1} (1 - P_+^p) + \sum_{q = 1}^{m_2} (1 - (1 - P_+^q))
 \end{equation}
where there are $m_1$ training samples labelled as ``malignant'' and $m_2$ samples labelled as ``benign''. See Appendix C for more details about obtaining the loss function in equation~\ref{error_lin_bin} from the cross-entropy loss function in equation~\ref{crossentropy} above.

\vspace{0.4em}
Fig.~\ref{ibmq_graph} (a) and (b) show the values of $P_+$ and $(1-P_+)$, which are generated by a trained variational quantum circuit for a particular ``malignant'' sample and a particular ``benign'' sample respectively, in the case of noiseless simulation on Qiskit (platform 2 above). Fig.~\ref{ibmq_graph} (c) and (d) show the values of $P_+$ and $(1-P_+)$ for the same ``malignant'' sample and the same ``benign'' sample respectively in the case of implementation on IBM-Q (platform 3 above). The contribution of noise in IBM-Q can be understood by comparing the plots for IBM-Q-based implementation with that for noiseless Qiskit simulation.  As expected, in the case of the Qiskit simulation, we note that the probability values $P_+$ and $(1-P_+)$ are closer to the ideal values ($P_+ \sim 1$ for ``malignant'' and $(1-P_+) \sim 1$ for ``benign'') compared to the IBM-Q-based implementation.

\vspace{0.4em}
For both Qiskit and IBM-Q implementation, the sample is classified as ``malignant'' if $P_+ > c_t$ and ``benign'' if $1-P_+ > c_t$. Noise in the IBM-Q system does not lead to wrong classification of data if $c_t$ has a reasonably low value. Fig.~\ref{ibmq_graph} (c) and (d) (results on one particular sample of each type) indicate that choosing $c_t$ within the range 0.5—0.6 may lead to successful classification of many such samples if similar probability-numbers are obtained for those other samples as in Fig.~\ref{ibmq_graph} (c) and (d).

\begin{table}[!t]
\begin{center}
\begin{tabular}{| l | l | l |}  
\hline
{$\mathbf{c_t}$} & {\bf Qiskit} & {\bf IBM-Q} \\ \hline
0.5 & 96.45$\%$ & 91.71$\%$ \\ \hline
0.6 & 94.08$\%$ & 87.57$\%$ \\ \hline
0.7 & 89.34$\%$ & 76.33$\%$ \\ \hline
0.8 & 80.47$\%$ & 62.72$\%$ \\ \hline
0.9 & 62.72$\%$ & 31.95$\%$ \\ \hline
\end{tabular}
\end{center}
\caption{\label{tab:ct} Classification accuracy numbers on the test data-set for WBC (test accuracy) for different values of classification-metric $c_t$ (defined in Section II) are tabulated here. We obtain the numbers after implementing the algorithm on Qiskit (platform 2) and  IBM-Q (platform 3).}
\end{table}

\vspace{0.4em}
Table ~\ref{tab:ct} agrees with that observation since it shows reasonably high classification accuracy on the entire test data-set for the IBM-Q platform for the range of $c_t$ mentioned above. Again, this shows that our model is robust against noise within a reasonable range of values for the classification-metric ($c_t$). When the value of $c_t$ is very high, $P_+$ and $(1-P_+)$ deviate from their ideal values to a large extent for the IBM-Q platform (since it has noise). So the classification accuracy drops drastically in Table ~\ref{tab:ct} for the IBM-Q-based implementation compared to the noiseless Qiskit-based implementation. 

\vspace{0.4em}
Even for the noiseless Qiskit-based implementation, the classification accuracy drops to a degree with increase in  $c_t$ (Table ~\ref{tab:ct}). This is because the loss in equation \ref{error_lin_bin} cannot be minimized to the lowest possible value for every sample in the data-set. This leads to some input-samples being classified wrongly when the classification criterion is stringent (high value of $c_t$), just like in any classical ML algorithm. We identify these input samples on the Bloch sphere, corresponding to our trained network, in Section IV next.

\section{Explanation of the Working of the Algorithm Using a Bloch Sphere}

\begin{figure*}[!t]
      \centering
      \includegraphics[width=0.8 \textwidth]{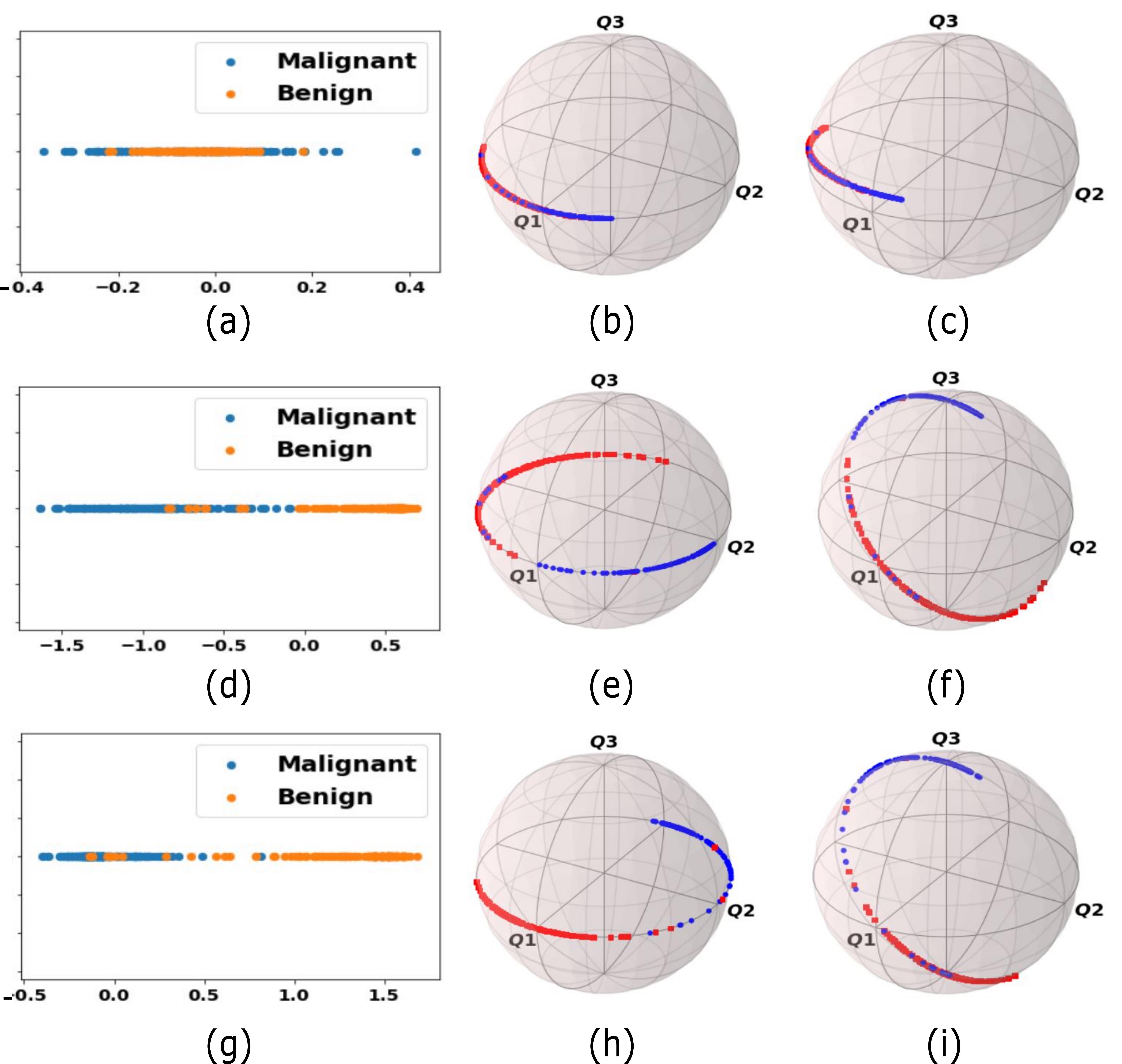}
      \caption{Training using the WBC data-set: (a), (d), and (g) show the output of the classical encoding layer (${\cal N}_{30 \rightarrow 1}$):  $\Tilde{x}_j = \sum_i x_i w_i^j$. (b), (e), and (h) show the states $\psi({\bf x})$ (equation ~\ref{eq:state_prep}) on the Bloch sphere, for all training samples ${\bf x}$. (c), (f), and (i) show the states $\vert \bar{\psi}({\bf x})$ \big> (equation ~\ref{eq:su2trans})  on the Bloch sphere, for all training samples. $Q_1$, $Q_2$, and $Q_3$ are the three components of the polarization vector, corresponding to the Bloch sphere. (a), (b), and (c) correspond to the end of the 1st epoch; (d), (e), and (f) correspond to the end of the 20th epoch; (g), (h), and (i) correspond to the end of the 200th epoch. A blue dot corresponds to a sample belonging to the class ``malignant''.  A red dot corresponds to a sample that belongs to the class ``benign''. }
      
      \label{fig:visual}
     \end{figure*}

\begin{figure*}[!t]
      \centering
      \includegraphics[width=0.8 \textwidth]{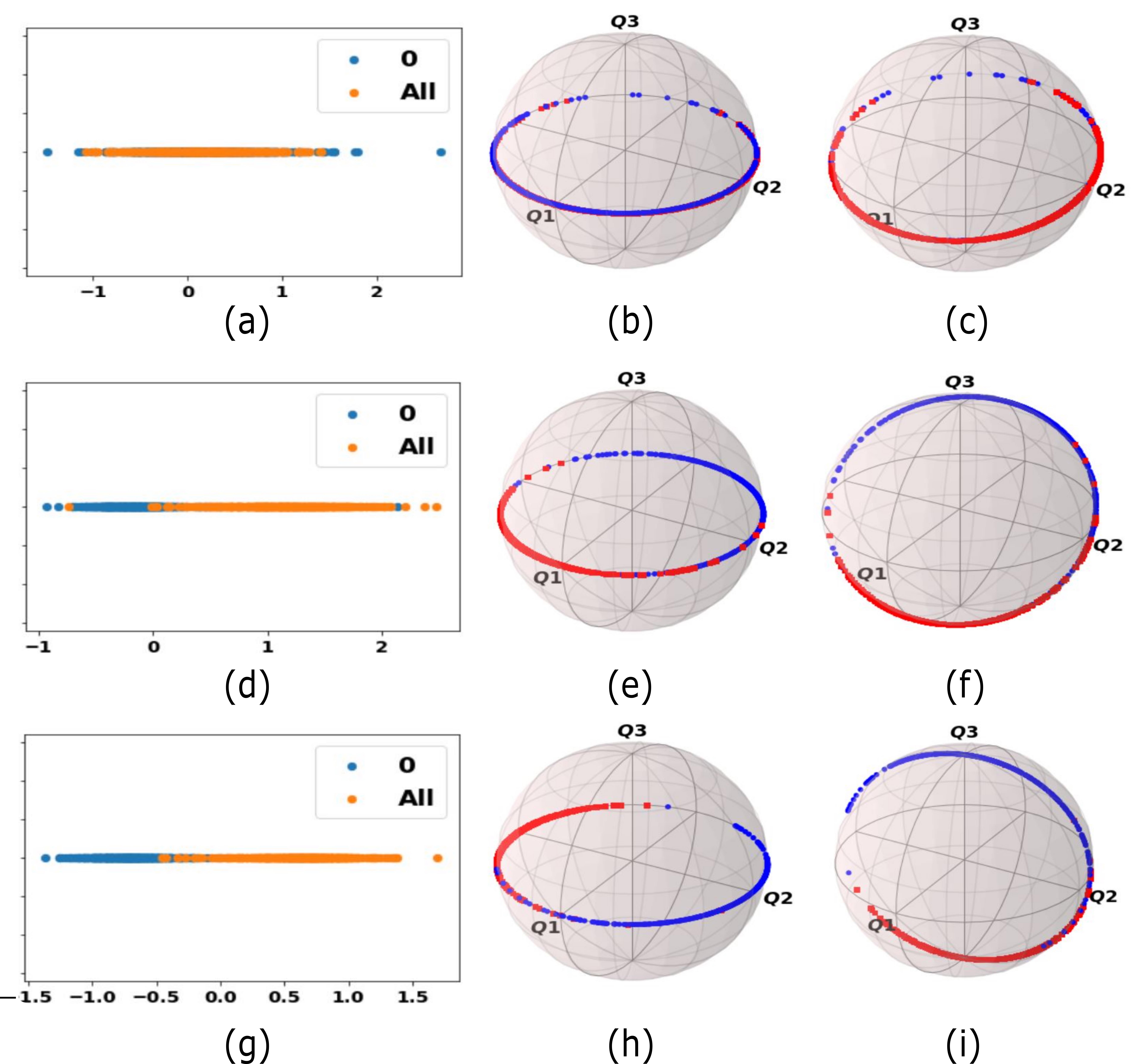}
      \caption{Training using the MNIST data-set (for binary classification: digit ``0'' vs. all other digits): (a), (d), and (g) show the output of the classical encoding layer (${\cal N}_{784 \rightarrow 1}$):  $\Tilde{x}_j = \sum_i x_i w_i^j$. (b), (e), and (h) show the states $\psi({\bf x})$ (equation ~\ref{eq:state_prep}) on the Bloch sphere, for all training samples ${\bf x}$. (c), (f), and (i) show the states $\vert \bar{\psi}({\bf x})$ \big> (equation ~\ref{eq:su2trans})  on the Bloch sphere, for all training samples. $Q_1$, $Q_2$, and $Q_3$ are the three components of the polarization vector, corresponding to the Bloch sphere. (a), (b), and (c) correspond to the end of the 1st epoch. (d), (e), and (f) correspond to the end of the 20th epoch. (g), (h), and (i) correspond to the end of the 200th epoch. A blue dot corresponds to a sample belonging to the class: digit ``0''.  A red dot corresponds to a sample that belongs to the class: all other digits.}
      
      \label{fig:visual_MNIST}
     \end{figure*}

\subsection{Bloch-Sphere-Based Representation for the Binary Classification Problem}

\vspace{0.4em}
To gain better insight into the working of our training algorithm and understand the precise role of the different parameters that we adjust iteratively during the training process, we represent the states in equation ~\ref{eq:state_prep} and equation \ref{eq:su2trans} on a Bloch sphere \cite{nielsen2002, blum2012}. We visualize their evolution during the training process over multiple epochs. Here, we have chosen the case of binary classification for the sake of simplicity. A binary classification problem requires only one qubit, as explained in Section III; this makes the Bloch-sphere-based visualization easy.

\vspace{0.4em}
Fig.~\ref{fig:visual} and Fig.~\ref{fig:visual_MNIST} show how our proposed dressed quantum network gets trained for the binary classification problem with the WBC dataset (2-class-data-set) and MNIST data-set of handwritten digits \cite{lecun-mnisthandwrittendigit-2010}. The MNIST data-set originally contains data for ten classes, corresponding to the digits: 0–9.  But we have reformulated the ten-class-classification problem to ten separate binary-classification problems of the type: digit ``$i$'' vs. all other digits ($i \in \{0, \cdots, 9\}$). Fig.~\ref{fig:visual_MNIST} shows the Bloch-sphere-based representation of the qubit during the training process for the problem: digit ``0'' vs. all other digits (in the MNIST data-set) \cite{lecun-mnisthandwrittendigit-2010}. We carefully choose the training data from the original MNIST data-set to avoid imbalance between training data corresponding to digit ``$i$'' and that corresponding to all other digits \cite{longadge2013class,abd2013review}.  For more details about how we use the MNIST data-set in our paper, see Appendix A below. The results shown here have been obtained by executing the classifier-algorithm on a classical computer using a Python-code (Platform 1, as listed in Section III). 



\vspace{0.4em}
The variational quantum circuit now contains a single qubit. The sequence of parametrized gates is still the same as before (Section II). Following our ``super compressed encoding'' scheme, the input vector is first transformed into a single number; ${\cal N}: {\bf x} \rightarrow \Tilde{x} = \sum_i x_i w_i$, $i  = 1, 2, \cdots, d$. For the WBC data-set, $d=30$; for the MNIST data-set, $d=784$.  The scalar $\Tilde{x}$ is then encoded into a single qubit state $\ket{ \psi (\bf x)} = e^{i \sigma_3 \Tilde{x}} H \ket{0}$. A parametrized SU(2) operation ($e^{i \sigma_3 \alpha_1} e^{i \sigma_2 \alpha_2} e^{i \sigma_3 \alpha_3}$)  is subsequently applied on $\vert \psi({\bf x})\big>$ to produce $\ket{\bar \psi (\bf x)}$. A projective measurement is then performed on the state and the probability for the outcome $\sigma_3 = +1$ is recorded as $P_+$. 
 
\vspace{0.4em}
The classification metric $c_t$ is chosen to be 0.5 here. So, for the WBC dataset, a sample is classified as ``malignant'' if $P_+ > 0.5$; it is classified as ``benign'' if $(1-P_+) > 0.5$. Similarly, for the ``0'' vs. all other digits problem (MNIST), a sample is classified as digit ``0'' if $P_+ > 0.5$; it is classified as ``other digit'' if $(1-P_+) > 0.5$.  The model is trained such that $P_+ \rightarrow 1$ for the samples labelled as ``malignant''/ digit ``0'' and $(1-P_+) \rightarrow 1$ for the samples that are ``benign''/ ``other digit''. Intuitively, this means that, as a part of the training process, the weights in the classical encoding layer and the rotation angles in the SU(2) operation are adjusted to ensure that $\ket{\bar \psi (\bf x)}$ evolves to $\ket{0}$ if ${\bf x}$ belongs to the class ``malignant''/ digit ``0'' while $\ket{\bar \psi (\bf x)}$ evolves to $\ket{1}$ if ${\bf x}$ belongs to the class ``benign''/ ``other digit'' (as mentioned earlier in Section III).

 \vspace{0.4em}
Fig.~\ref{fig:visual} and Fig.~\ref{fig:visual_MNIST} depict the evolution of the following quantities, over several epochs, during the training process:

\begin{enumerate}
    \item The  output of the classical encoding layer for all training samples:  Fig.~\ref{fig:visual} (a), (d), and (g) show the quantity  ${\Tilde{x}}$ at the end of the 1st, 20th, and the 200th epoch respectively, for the WBC data-set. Fig.~\ref{fig:visual_MNIST} (a), (d), and (g) show the quantity  ${\Tilde{x}}$ at the end of the 1st, 20th, and the 200th epoch respectively, for the MNIST data-set (digit ``0'' vs. all other digits). 
    
    \item The state $\ket{\psi (\bf x)}$  on the surface of a Bloch  sphere for all training samples. Fig.~\ref{fig:visual} (b), (e), and (h) show the state $\ket{\psi (\bf x)}$ at the end of the 1st, the 20th, and the 200th epoch respectively, for the WBC data-set. Fig.~\ref{fig:visual_MNIST} (b), (e), and (h) show the state $\ket{\psi (\bf x)}$ at the end of the 1st, the 20th, and the 200th epoch respectively, for the MNIST data-set (digit ``0'' vs. all other digits). 
    
    \item The state $\ket{\bar \psi (\bf x)}$  on the surface of a Bloch  sphere for all training samples. Fig.~\ref{fig:visual} (c), (f), and (i) show the state $\ket{\bar \psi (\bf x)}$ at the end of the 1st, the 20th, and the 200th epoch respectively, for the WBC data-set. Fig.~\ref{fig:visual_MNIST} (c), (f), and (i) show the state $\ket{\bar \psi (\bf x)}$ at the end of the 1st, the 20th, and the 200th epoch respectively, for the MNIST data-set (digit ``0'' vs. all other digits).
\end{enumerate}

\vspace{0.4em}
 Any single-qubit state \(\ket{\psi}\) can be written as a density matrix of the form $\rho = \ket{\psi}\bra{\psi}  = \frac{1}{2} ({\bf 1} + {\bf Q}\cdot {\bf \sigma})$; with ${\bf Q}$ being the polarization vector in three dimensions with $\vert \vert {\bf Q} \vert \vert_{L_{2}} \leq 1$ \cite{nielsen2002,blum2012}. This polarization vector defines a single qubit state uniquely. The Bloch sphere is a unit sphere in the three-dimensional space, defined by the components of the polarization vector \textbf{Q} which are: $Q_1$, $Q_2$, and $Q_3$. Each point on the surface of the Bloch sphere represents a unique ${\bf Q}$, which corresponds to a unique pure state \(\ket{\psi}\). The probability of the outcome $\sigma_3 = +1$, for the state $\rho$ is given by the overlap of the density matrix with the projection operator $\pi_+ = \frac{1}{2} ({\bf 1} + \sigma_3)$:
\begin{equation}
 \label{Bloch1}
    P_+ = Tr(\rho \pi_+) = \frac{1}{2}(1 + Q_3)
\end{equation}

Any state $\ket{\bar\psi(\bf x)}=e^{i \sigma_3 \alpha_1} e^{i \sigma_2 \alpha_2} e^{i \sigma_3 \alpha_3}  e^{i \sigma_3 \Tilde{x}} H(\ket{0})$ (where $\bf x$ = $\{x_1,x_2,x_3,...x_d\}$ and $\Tilde{x} = \sum_i x_i w_i$, $i  = 1, 2, \cdots, d$) can be represented on the surface of the Bloch sphere as ${\bf Q}$= $\{Q_1, Q_2, Q_3\}$ through the following mapping:

\begin{eqnarray}
 \label{Bloch2}
&Q_1& = \cos^2\alpha_2 \cos 2 (\alpha_1+\alpha_3+\Tilde{x})-\sin^2 \alpha_2 \cos 2 (\alpha_1-\alpha_3-\Tilde{x}) \nonumber \\
&Q_2& =  \sin^2\alpha_2 \sin 2 (\alpha_1-\alpha_3- \Tilde{x})-\cos^2\alpha_2 \sin 2 (\alpha_1+\alpha_3+ \Tilde{x}) \nonumber \\
&Q_3& = \sin 2 \alpha_2 \cos 2(\alpha_3 + \Tilde{x})
\end{eqnarray}

\vspace{0.4em}
The blue dots on the Bloch sphere in Fig.~\ref{fig:visual} correspond to the ``malignant'' samples while the red dots represent the samples belonging to the ``benign'' class. The blue dots in Fig.~\ref{fig:visual_MNIST} correspond to the digit ``0'' samples while the red dots represent the samples belonging to the class: all other digits/ ``other digits''. Indeed, just as we desired,  Fig.~\ref{fig:visual} (c), (f) and (i) and Fig.~\ref{fig:visual_MNIST} (c), (f) and (i) show that after training the circuit over a suitable number of epochs, the network learns how to differentiate between the samples that are labelled differently. For ``malignant'' samples/ digit ``0'' samples (blue dots), the corresponding states \(\ket{\bar \psi(\bf x)}\) are adjusted such that \(\ket{\bar \psi(\bf x)}\) evolve towards $\ket{0}$. So the third component of the polarization vector (${Q}_3$) is positive. Thus, all such training samples lie on the upper hemisphere of the Bloch sphere thereby ensuring the condition $P_+ > 0.5$ (from equation ~\ref{Bloch1}) . Similarly, for the ``benign'' samples/ ``other digit'' samples, \(\ket{\bar \psi(\bf x)}\) is adjusted such that  \(\ket{\bar \psi(\bf x)}\) evolves towards $\ket{1}$. These points lie on the lower hemisphere of the Bloch sphere (${Q}_3 < 0$, and hence from equation ~\ref{Bloch1},  $(1 - P_+) > 0.5$). 

\vspace{0.4em}
It is to be noted that while we do try to train our dressed network such that ${Q}_3$ for all ``malignant''/ digit ``0'' samples attains the highest possible positive value ($\to 1$) and  ${Q}_3$ for all ``benign''/ other digit samples attains the lowest possible negative value ($\to -1$), this is not possible for all the samples. This is because the loss-term in equation ~\ref{error_lin_bin} cannot be minimized to the lowest possible value for each and every sample. We observe this phenomenon in Fig.~\ref{fig:visual} (i) and Fig.~\ref{fig:visual_MNIST} (i). For both the clusters, corresponding to the two output-classes, there are many points on the Bloch sphere further away from the ideal points: ${Q}_3 = 1$ (for ``malignant''/ digit ``0'' samples) and ${Q}_3 = -1$ (for ``benign''/ other digit samples). These samples here are taken from the training data-set. When \(\ket{\bar \psi( \bf x)}\) corresponding to these test-samples is plotted on the Bloch sphere with parameters in the classical encoding layer and rotation parameters in SU(2) operations being the ones obtained after training, a similar trend will be observed. As the classification-metric $c_t$ (in Section II) assumes a higher value, the samples corresponding to these points tend to get wrongly classified, leading to a drop in the test accuracy (Table ~\ref{tab:ct}), as discussed before in Section III.

\vspace{0.4em}
Additionally, in Table ~\ref{table_mnist}, we report the test classification accuracy on the test data-set for this binary classification problem with MNIST (digit ``0'' vs. all other
digits, digit ``1'' vs. other digits, etc.). We observe that in all these cases, the classification accuracy is fairly high. This shows that our ``super-compressed encoding'' scheme is still efficient even for high-dimensional input (784 for MNIST).

\begin{table}[!t]
\begin{center}
\begin{tabular}{| l | l |} 
\hline
{\bf Dataset} & {\bf Accuracy}   \\ \hline
MNIST (0 vs All) &  97.80 $\%$  \\ \hline
MNIST (1 vs All) &  97.22 $\%$  \\ \hline
MNIST (2 vs All) &  91.61 $\%$  \\ \hline
MNIST (3 vs All) &  91.58 $\%$  \\ \hline
MNIST (4 vs All) &  92.97 $\%$  \\ \hline
MNIST (5 vs All) &  90.91 $\%$  \\ \hline
MNIST (6 vs All) &  95.45 $\%$  \\ \hline
MNIST (7 vs All) &  94.01 $\%$  \\ \hline
MNIST (8 vs All) &  88.60 $\%$  \\ \hline
MNIST (9 vs All) &  90.08 $\%$  \\ \hline
\end{tabular}
\end{center}
\caption{\label{table_mnist} Classification accuracy numbers for the binary classification problem of MNIST as obtained by running our algorithm on a classical computer using Python code (platform 1). Value of the classification metric $c_t$ is chosen to be 0.5 here.}
\end{table}

\subsection{Distinct Roles Played in Training by the Classical Encoding Layer and the SU(2) Operations}

\vspace{0.4em}
Fig. ~\ref{fig:visual} and Fig. ~\ref{fig:visual_MNIST}  reveal the distinct role, with respect to the binary classification above, that has been played by the two sets of iteratively adjustable parameters: the weights in the classical encoding layer ($w_i$; $i=1,2,3,...d$) and the rotation/ SU(2) parameters ($\alpha_1,\alpha_2,\alpha_3$) in the variational quantum circuit. The weights in the classical encoding layer are updated, after every epoch, during training such that after the training (say the 200th epoch), for input-samples ($\bf x$) belonging to ``malignant'' type/ digit ``0'', the corresponding $\ket{\psi(\bf x)}$-s cluster on one part of the circle on the Bloch sphere with $Q_3=0$; for  input-samples belonging to ``benign'' type/ all other digits, $\ket{\psi(\bf x)}$-s cluster on the opposite part of the circle on the Bloch sphere with $Q_3=0$. This can be seen from Fig. \ref{fig:visual}(h) and Fig. \ref{fig:visual_MNIST}(h).

\vspace{0.4em}
But we also observe from Fig. ~\ref{fig:visual}(h) and Fig. ~\ref{fig:visual_MNIST}(h) that though $\ket{\psi(\bf x)}$-s corresponding to the samples belonging to the two output-classes are well separated on the circle ($Q_3=0$), the axis that separates these two clusters cannot be determined with certainty. For example, the location of the two clusters formed corresponding to the two output-classes is different for the WBC data-set (Fig. ~\ref{fig:visual}(h)) and the MNIST data-set (Fig. ~\ref{fig:visual_MNIST}(h)). Thus adjusting the weights in the classical encoding layer is not enough to identify, with certainty, which sample belongs to which output-class. The SU(2) parameters in the variational quantum circuit also need to play an important role in this identification, as we explain next.

\vspace{0.4em}
These SU(2) parameters ($\alpha_1,\alpha_2,\alpha_3$) are learned during training such that after training (after the 200th epoch say), for one sample-type (``malignant''/digit ``0''), $\ket{\psi(\bf x)}$, after the SU(2) operations, transforms to $\ket{\bar \psi(\bf x)}$, which evolves towards $\ket{0}$ (Fig.~\ref{fig:visual}(i), Fig.~\ref{fig:visual_MNIST}(i)). Thus all the $\ket{\bar \psi(\bf x)}$-s for such samples cluster towards the point on the Bloch sphere with $Q_3=1$. Hence, for such samples, ${Q}_3$ is positive and $P_+ > 0.5$ as mentioned earlier. Similarly, for the other sample-type (``benign''/other digits), $\ket{\psi(\bf x)}$, after the SU(2) operations, transforms to $\ket{\bar \psi(\bf x)}$, which evolves towards $\ket{1}$ (Fig.~\ref{fig:visual}(i), Fig.~\ref{fig:visual_MNIST}(i)). Thus all the $\ket{\bar\psi(\bf x)}$-s for such samples cluster towards the point on the Bloch sphere with $Q_3=-1$. Hence, for such samples, ${Q}_3$ is negative and $(1 - P_+) > 0.5$ as mentioned earlier. The minimization of the loss function in equation ~\ref{error_lin_bin} enables such SU(2)-parameter adjusting.

\vspace{0.4em}
Overall, from Fig.~\ref{fig:visual} and Fig.~\ref{fig:visual_MNIST}, we observe that updating the weight parameters in the classical encoding layer enables the clustering of the samples based on their output-classes. But the clusters are formed on different randomly located parts of the circle on the Bloch sphere with ${Q}_3=0$. The iterative adjustment of the rotation parameters in the SU(2) operation enables driving these two randomly formed clusters towards specific points on the Bloch sphere:  $Q_3=1$ and $Q_3=-1$; this allows identifying each sample as being of one output-class or another and leads to high classification accuracy. 

\vspace{0.4em}
It is to be noted that $P_+$ (the outcome of a measurement on the quantum state) in equation ~\ref{error_lin_bin} not only depends on the SU(2) parameters in the quantum circuit but also the weight parameters in the classical encoding layer. These weight parameters are also iteratively adjusted to minimize the loss function in equation ~\ref{error_lin_bin}. So the training of the classical encoding layer also indirectly depends upon the SU(2) operations in the quantum circuit, the measurement process, and the iterative adjustment of the SU(2) parameters during the training of the quantum circuit. Thus, training of the classical encoding layer and training of the quantum circuit depend on each other. Together, they lead to the training of the overall dressed quantum network.


\section{Advantages of Our Algorithm}

\vspace{0.4em}
We now compare the data-sets that we have classified, with high accuracy, using our proposed QML algorithm (Table ~\ref{tab:accuracy}, ~\ref{table_mnist}) with that used in existing QML algorithms. The precise data-sets classified by these algorithms indicate the kind of ML-tasks they are capable of solving. Once we show that our algorithm can classify the data in data-sets as complex or more complex than the ones used by existing QML algorithms, we will show the explicit advantages that our algorithm offers.

\vspace{0.4em}
Table ~\ref{table_datasetcomp} shows that for many existing QML algorithms, classification has only been reported only for toy data-sets, constructed for this purpose in the reports on these algorithms themselves \cite{salinas2020,havlivcek2019,Tachino}. On the contrary, in this paper, we report high classification accuracy on four standard ML data-sets: Fisher's Iris \cite{dua2019uci}, WBC \cite{dua2019uci}, Abalone \cite{dua2019uci}, and MNIST \cite{lecun-mnisthandwrittendigit-2010}. 
While \cite{OxfordXORIris} and  \cite{MariaIris} also show classification using Fisher's Iris data-set, we also show classification, as mentioned here, using Abalone and MNIST data-set. If we judge the complexity of a data-set by the number of input-dimensions, MNIST and WBC data-set have much higher input-dimensions (784 and 30 respectively) compared to Fisher's Iris data-set (4) and most toy data-sets used in \cite{salinas2020,havlivcek2019,Tachino}.  If we judge the complexity of a data-set by the number of output-classes, Abalone data-set has more output-classes (6) than Iris (3) and most toy data-sets used in \cite{salinas2020,havlivcek2019,Tachino}.

\vspace{0.4em}
Thus we show through Table ~\ref{table_datasetcomp} that our proposed QML algorithm can classify ML data-sets as complex as or more complex than the ones used by different existing QML algorithms. We can achieve this result despite using a ``super compressed encoding'' scheme and drastically scaling down the input dimensions to just one scalar per input-sample. To the best of our knowledge, only in \cite{mari2019}, the data-sets used are more complex than ours. However, they use a ResNET-block to extract the essential features from the input-images \cite{ResNet}. Since ResNET is a very complex neural network classically pre-trained on the ImageNet data-set, it learns the complex structure of images. It classically produces only those features that are essential for the final classification-task \cite{ImageNet}. On the contrary, in this paper, we initialize all the parameters of our classical encoding layer to random values. Then we iteratively adjust them to final values during the training process, which also involves iteratively adjusting the SU(2) parameters of the variational quantum circuit.

\begin{table*}[!t]
\begin{center}
\begin{tabular}{| l | l |} 
\hline
{\bf Reference of the QML algorithm} & {\bf Data-sets used}   \\ \hline
A. Perez-Salinas \textit{et al.}, Quantum \textbf{4}, 226 (2020) \cite{salinas2020} &  Toy data-set constructed in \cite{salinas2020} to classify \\
  & if different points lie inside or outside specific geometric regions  \\
  \hline
V. Havlickek \textit{et al.}, Nature \textbf{567}, 209 (2019) \cite{havlivcek2019} & Toy data-set constructed in \cite{havlivcek2019} to classify different kinds of points \\
\hline
F. Tacchino \textit{et al.},npj Quantum Information \textbf{5}, 26 (2019) \cite{Tachino} & Toy data-set constructed in \cite{Tachino} consisting of black-and-white patterns \\
  \hline
D. Zhu \textit{et al.}, Science Advances \textbf{5}, eaaw9918 (2019) \cite{CQC2019} & Bars-and-Stripes data-set \cite{BarsStripes} \\
\hline
M. Benedetti \textit{et al.}, npj Quantum Information \textbf{5}, 1 (2019) \cite{benedetti2019} & Bars-and-Stripes data-set \cite{BarsStripes} \\
\hline
M. Schuld \textit{et al.}, Europhysics Letters \textbf{119}, 6 (2017) \cite{MariaIris} & Fisher's Iris data-set \cite{dua2019uci} \\
\hline
S. Cao \textit{et al.}, Physical Review A \textbf{101}, 052309 (2020) \cite{OxfordXORIris} & classical XOR gate, 
Fisher's Iris data-set \cite{dua2019uci} \\
\hline
A. Mari \textit{et al.}, arxiv: 1912.08278 (2019) \cite{mari2019} & ImageNet, CIFAR-10 \cite{mari2019} \\
\hline
This paper & Fisher's Iris \cite{dua2019uci}, WBC \cite{dua2019uci}, Abalone \cite{dua2019uci}, MNIST \cite{lecun-mnisthandwrittendigit-2010} \\
\hline
\end{tabular}
\end{center}
\caption{\label{table_datasetcomp} Data-sets used by the different existing QML algorithms for classification}
\end{table*}

\vspace{0.4em}
Having shown that our proposed QML algorithm can solve ML-tasks as complex or more complex than most existing QML algorithms, we now discuss the different advantages of our proposed algorithm compared to the existing algorithms. We had already mentioned a couple of advantages in Section I: robustness against noise, and a low number of qubits. Here, we use quantitative estimates to compare our proposed algorithm with other existing QML algorithms for these two metrics. We also discuss some other advantages of our proposed algorithm below. 

\begin{enumerate}
  \item \textbf{Robustness against noise:} 
  As mentioned in Section I, multi-qubit gates are absent in our algorithm. We only use single-qubit gates; this leads to low noise in our implementation (Table ~\ref{tab:ct}, Fig. ~\ref{ibmq_graph}). All multi-qubit gates, when expressed in the basis-gate-set of IBM-Q devices, require the implementation of CNOT gates. On IBM-Q devices, the error-rate of CNOT gates is 1–2 orders of magnitude higher than that of the single-qubit gates. For example, on IBM-Q Melbourne, the CNOT error-rate ranges from $1.33\%$ to $6.18\%$. But for the single-qubit U2 gate, the error-rate ranges from $0.038\%$ to $0.412\%$\cite{ibmq}. Although the exact numbers change frequently, the approximate ratio of noise in the multi-qubit gate to the single-qubit gate almost remains the same, as seen here. Thus using multi-qubit gates makes implementing the algorithm more error-prone on current NISQ quantum devices.
   
  \item \textbf{Low number of qubits and quantum gates:} As mentioned in Section I, the ``super compressed encoding'' scheme used in this paper enables us to reduce the number of qubits in the variational quantum circuit compared to existing QML algorithms \cite{havlivcek2019}. The number of qubits in our implementation is independent of the input-dimensions and only depends on the number of output-classes. Hence, for the WBC data-set used above (2 output-classes), we use can only two qubits to encode each input-sample. Since this is a binary classification problem, we can further modify our algorithm to use only one qubit for that purpose (Fig. \ref{fig:visual}). On the contrary, in \cite{havlivcek2019}, 30 qubits will be needed to encode the same 30-dimensional input-sample in the WBC data-set.  
   
  \item \textbf{Implementation of the non-linear activation function:} Another advantage of our algorithm lies in how easily we can apply a non-linearity in our classifier. Most classifiers, classical or quantum, require applying a non-linear activation function after a linear layer. For deep neural networks, these non-linearities are generally ``sigmoid'' or ``tan-sigmoid'' functions \cite{lecun2015}. Evaluating such a function requires evaluating an exponential function as an intermediate step, which is an expensive operation on a digital classical computer. In classical neuromorphic computing (implementation of ML algorithms through unconventional, but classical, architectures and devices), such non-linear activation function is often implemented through transistor-based analog circuits (not digital CMOS circuits) or different other emerging devices \cite{Bhowmik_JMMM,TransistorSynapseBioCAS,PCMReview_AbuSebastian,GeffBurr_Nature_TransistorPCM,Kaushik_IEEEReview}. Here, in our implementation of an ML algorithm on quantum hardware (it can be called quantum neuromorphic computing \cite{Grollier_QNeuro2020}), our quantum operations in the variational quantum circuit provide us with that non-linear activation function \cite{Tachino}. For example, the quantum state preparation (\(e^{i \sigma_3 \Tilde{x}}\)) is non-linear with respect to the input that the classical encoding layer feeds to the system (\(\Tilde{x}\)). Similarly, the measurement of the qubit provides another non-linearity. 
     
  \item \textbf{Adaptability of the activation function:} The iteratively adjustable rotation parameters in the SU(2) operations on the qubits introduce adaptability in the activation function. From that perspective, our proposed algorithm can be compared with an algorithm recently used in classical neuromorphic computing where the input-data have been clustered through a linear network and oscillator-functions (their properties are adaptable) are used to learn the boundaries between the clusters \cite{Grollier_Nature2018}. Similarly in our algorithm, as explained in details in Section IV through the Bloch-sphere-based representation, iterative adjustment of the weights in the classical encoding layer enables separating the data into different clusters on the circle of the Bloch sphere with $Q_3 = 0$ (Fig. ~\ref{fig:visual} (h), Fig. ~\ref{fig:visual_MNIST} (h)). Iterative adjustment of the SU(2) parameters in the variational quantum circuit maps these clusters, formed at random positions on that circle on the Bloch sphere, to specific parts of the Bloch sphere as shown in Fig. ~\ref{fig:visual} (i) and Fig. ~\ref{fig:visual_MNIST} (i). Such adaptive property of activation functions has been found to be very useful for data-classification \cite{Grollier_Nature2018,AdaptiveNeuron1,AdaptiveNeuron2}.

  \item \textbf{Explainability:} As mentioned in Section I, to the best of our knowledge, such Bloch-sphere-based approach has not been used earlier to explain the working of other existing QML algorithms. The fact that we can explain the internal mechanism based on which our proposed QML algorithm works here (Section IV) is very helpful, given that so much research has been carried out recently to explain the internal mechanism of ML algorithms \cite{ExplainableAI1,ExplainableAI2}.

\end{enumerate}

\section{Conclusion}

\vspace{0.4em}
In this paper, we have proposed and implemented a QML algorithm that uses a dressed quantum network. The classical encoding layer in our dressed network uses the ``super compressed encoding'' scheme to drastically scale down the input-dimensions. We use the Bloch-sphere-based representation to explain the working of our algorithm. We implement our algorithm on a classical computer, using Python code, as well as on Qiskit and real NISQ hardware (IBM-Q). We report high classification accuracy numbers for our implementation on different ML data-sets. We show that our algorithm can handle ML data-sets of the complexity of the ones that other existing QML algorithms typically deal with. We also argue that our algorithm has several advantages compared to various existing QML algorithms: a low number of qubits, robustness against noise, implementation of adaptable non-linear activation functions, etc.

\begin{acknowledgments}
Debanjan Bhowmik thanks Department of Science and Technology (DST), India, for the INSPIRE Faculty Award, and Science and Engineering Research Board (SERB), India, for the Early Career Research (ECR) Award. Debanjan Bhowmik also thanks Industrial Research and Development Unit, Indian Institute of Technology Delhi and Nokia Networks, India, for the Discover and Learn 1-2-3-4 project. These awards and projects  funded this research. The authors also thank Soumik Adhikary (from Department of Physics, Indian Institute of Technology Delhi) and Rajamani Vijayaraghavan (from Tata Institute of Fundamental Research, Mumbai, India) for their insights about our proposed algorithm.
\end{acknowledgments}

\appendix

\section{The Machine Learning Data-Sets Used Here}

\subsection*{Fisher's Iris Data-Set}
\vspace{0.4em}
Fisher's Iris dataset has 150 samples in all; there are 50 samples each for the three Iris flower-species: ``setosa'', ``virginica'', and ``versicolor'' \cite{dua2019uci}. Each input-sample has 4 features: septal-length, septal-width, petal-length, and petal-width. The task is to correctly categorize a given sample into one of the three classes representing the three flower-species.


\vspace{0.4em}
To accomplish the classification task, our algorithm needs to train a total of 21 parameters. 12 of these 21 parameters are the weights in the classical encoding layer (4 input-dimensions and 3 output-classes/ qubits). The rest are the rotation parameters in the SU(2) operations/ SU(2) parameters. The classification accuracy for this data-set is obtained for all the three platforms, mentioned in the paper (see Table \ref{tab:accuracy}). While running the algorithm on the classical computer by solving equations ~\ref{eq:state_prep} - \ref{crossentropy} (Platform 1), the data-set has been divided into a training set of 120 samples (40 samples per class) and a test-set of 30 samples (10 samples per class). The model has been trained over 20 epochs. We calculate the gradients using first-order numerical differentiation and minimize the cross-entropy loss function using the Adam optimizer \cite{kingma2014adam}.

\vspace{0.4em}
While implementing the algorithm on Qiskit (Platform, 2) and IBM-Q (Platform 3), we make slight variations to the training conditions. For execution on both Qiskit and IBM-Q, the Penny Lane package has been used for gradient-calculation and loss-minimization \cite{pennylane2018}. The cross-entropy loss has been minimized using the RMS-prop optimizer. The error has been minimized over 12 epochs. We have used the IBM-Q Rome machine to implement our algorithm on this data-set. The implementation requires three qubits; IBMQ-Rome, being a 5-qubit machine, is suitable for the purpose.

\subsection*{WBC Data-Set}
\vspace{0.4em}
The Wisconsin (Diagnosis) Breast cancer (WBC) data-set has 569 samples in all. Each sample has 30 real-valued input features, a unique ID (which we ignore for training), and a single label \cite{dua2019uci} . The samples can either be labeled as ``malignant'' or ``benign''. Each sample represents the features of the Fine Needle Aspirate (FNA) of a breast mass in the digitized image. The task is to correctly classify an unknown sample as ``malignant'' or ``benign''.

\vspace{0.4em}
We treat this as a binary classification and hence use only a single qubit (see Section III). So the total number of trainable parameters used is 33: 30 training weights in the classical encoding layer and 3 SU(2) parameters.  Just like for Fisher's Iris data-set, classification accuracy for the WBC data-set has been computed on all the three platforms, mentioned above. When the algorithm has been run on a classical computer by solving equations ~\ref{eq:state_prep} – \ref{crossentropy} (Platform 1: Python code), the data-set has been split into a training set of 400 samples and a test-set of 169 samples. Similar to Fisher's Iris data-set, the gradients are calculated using first order numerical differentiation, and the cross-entropy loss function is minimized over 200 epochs using the Adam optimizer \cite{kingma2014adam}.

\vspace{0.4em}
Similar to the case for Fisher's Iris data-set, while running the algorithm on Qiskit (Platform 2) and IBM-Q (Platform 3), the PennyLane software package has been used \cite{pennylane2018} and the loss in equation \ref{error_lin_bin} has been minimized over 12 epochs, using the RMS-Prop optimizer. We have used the IBM-Q Armonk machine to implement our algorithm for this data-set. The implementation requires just one qubit. So a single-qubit device like IBMQ-Armonk has been used.

\subsection*{Abalone Data-Set}
\vspace{0.4em}
Abalone data-set has a total 4177 samples with 8 input features each \cite{dua2019uci}. The age of an Abalone shell can be calculated by counting the number of rings in the abalone shell and adding 1.5 to it. The task here is about predicting the age of the Abalone shell using the 8 input features. In the data-set, the number of rings varies from 1 to 29, and hence, the age varies from 2.5 to 30.5 years. To solve this as a classification task, we club the samples that belong to a given range of ring-numbers into a single output-class. For instance, we club samples having 1–5 rings as class one, having 6–10 rings as second class, having 11–15 rings as third class, and so on. In total, we get six output-classes since the number of rings on shell varies from 1 to 29.

\vspace{0.4em}
We solve this problem as a 6-class classification problem. Total 66 parameters are used; for each class, 8 weight-parameters in the classical encoding layer and 3 SU(2) parameters are used; so 11 parameters are used per class, and there are 6 output-classes. We use 67\% of the total data-set for training and 33\% for testing, i.e., we use 2797 samples for training and 1380 samples for testing. Training and testing data points are selected randomly from the total data-set. We compute the classification accuracy on all three platforms for Abalone data-set (see Table- \ref{tab:accuracy}). While implementing our algorithm on a classical computer, here also we evaluate equation ~\ref{eq:state_prep} –\ref{crossentropy} using Python code, calculate the gradients using first-order numerical differentiation, and minimize the cross-entropy loss function using the Adam optimizer \cite{kingma2014adam} over 12 epochs.

\vspace{0.4em}
While implementing the same on Qiskit (Platform 2) and IBM-Q (Platform 3), only slight modification is made. Similar to WBC and Fisher's Iris, we use the PennyLane software package to compute numerical gradients and optimize the cross-entropy loss function over 12 epochs using RMS prop optimizer. We have used the IBMQ Melbourne machine to implement our algorithm for this data-set. The Abalone data-set requires 6 qubits; the only publicly available IBM-Q backend, which can run this task, is IBMQ-Melbourne.

\subsection*{MNIST Data-Set}
\vspace{0.4em}
MNIST is a popular data-set consisting of grey-scale ($28 \times 28$)-pixels-images of handwritten digits between ``0'' and ``9'' \cite{lecun-mnisthandwrittendigit-2010}. The original data-set consists of 60,000 training images and 10,000 testing images. The training as well as testing data-set is approximately uniformly distributed among all the 10 ouput-classes (10 digits). Each class has approximately 6,000 images for training (some classes have slightly more images than others, e.g., class ``0'' has 5923 samples while class ``6'' has 5918) and approximately 1,000 images for testing (again some classes have slightly more images than others).

\vspace{0.4em}
In this paper, we have considered ten binary-classification problems in MNIST where we distinguish digit ``$i$'' from all other digits; $i \in \{0, 1, \cdots, 9\}$. The target is to train our model to correctly infer whether a given image is that of the digit ``$i$'' or not. We have accomplished this task by implementing the modified algorithm, discussed earlier, on platform 1 (classical computer, using a Python code).

\vspace{0.4em}
The train and the test data-sets are carefully chosen since we carry out binary classification here (as mentioned in Section IV) \cite{longadge2013class,abd2013review}. For example, while training the network to distinguish between digit ``$i$'' and all other digits, if we use all the images available in the training data-set, the training data for binary classification will have approximately 1 : 9 imbalance.  Since the images are distributed uniformly among all the classes in the original data-set, we will have nine times more images corresponding to other digits compared to digit ``$i$''. To prevent this from happening, while training for the problem: digit ``$i$'' vs. all other digits, we select images from the original data-set such that we have an equal number of images for digit ``$i$'' and for all other digits. Thus the number of images corresponding to each digit among all the other digits is about one-ninth of the number of images corresponding to digit ``$i$''.

\vspace{0.4em}
For each binary classification using MNIST, 787 parameters are used. 784 of these parameters are the weights in the classical encoding layer, while the remaining three are the SU(2) parameters in the variational quantum circuit. The model has been trained by minimizing the loss function in equation ~\ref{error_lin_bin} over 200 epochs using the Adam optimizer \cite{kingma2014adam}. The gradient-calculation has been performed by the first-order numerical approximation method.


\section{Implementation of the Algorithm on IBM-Q}

\vspace{0.4em}
We would ideally like to implement our entire training algorithm on quantum hardware. However, for publicly available IBM-Q backends, the queuing times and other overheads make the training process extremely time-consuming. Another important bottleneck, here, is the gradient-calculation required for cost minimization. The gradients are calculated numerically. Hence, getting the value of the gradient directly from the quantum hardware will require the samples to be passed through the circuit multiple times. Keeping these shortcomings in mind, we have designed an alternate method to implement the algorithm on quantum hardware. The process has been briefly described earlier in the paper. Here, we give a detailed account of the implementation. 

\vspace{0.4em}
We start by creating a noise model of an IBM-Q  device on Qiskit using the device-characterization data provided by IBM \cite{ibmq}. We perform a noisy simulation of our algorithm on Qiskit. The results generated in the process are likely to be similar to those generated in the real IBM-Q device. To ensure this similarity, we record the updated model-parameters after every training epoch. After the parameters have been recorded after every epoch, we use them to implement the parametrized classifier circuit on real quantum hardware as well as on a noisy Qiskit simulator, for a fixed number of randomly sampled training data points. 

\vspace{0.4em}
It is to be noted that we perform just the forward pass here. The outputs of the classifier circuits are noted for both the platforms. This leaves us with two distributions D1 (from IBM-Q) and D2 (from noisy Qiskit simulator) for each randomly sampled training data point at the end of every epoch. The similarity between the two distributions is ensured by the condition $H(D_1, D_2)< \tau_1$. $H(D_1, D_2)$ is the Hellinger distance between the two distributions and $\tau_1$ is a threshold \cite{Hellinger}. Smaller the value of $\tau_1$, more is the similarity between the two distributions. We consider yet another threshold $\tau_2$ to make our comparison more stringent. $\tau_2$ is the minimum number of randomly chosen training samples that are used for the comparison. We expect the value of $\tau_2$ to be suitably large for a robust comparison. We find that the value of $\tau_1$ never exceeds $\sim 10^{-3}$  in our implementation. We have maintained a suitably high value of $\tau_2$ for all data-sets (\(\sim\) 10 \% of the total number of samples in the data-set). This ensures the similarity between the distributions obtained from the two platforms. This, in turn, ensures that implementing our algorithm on a noisy Qiskit simulator is equivalent to implementing the algorithm on a real quantum hardware.

\section{Obtaining the Loss Function for the Binary Classification Problem}

\vspace{0.4em}

According to the algorithm used in Section II and Section III, for a 2-class (binary) classification problem,  2 qubits are necessary in the variational quantum circuit. Say there are two classes, labelled 1 and 2 (e.g. ``malignant'' and ``benign'' in the case of the WBC data-set). There are $m_1$ samples of the class labelled 1 and $m_2$ samples of the class labelled 2. Using equation ~\ref{target} and equation ~\ref{crossentropy}, the net cross-entropy loss for all the samples can be expressed as:

\begin{equation}
\label{Derivation1}
\begin{split}
{\cal E}_{CrossEntropy} =    - \sum_{p = 1}^{m_1} log_{e} (\frac{e^{P^{1,p}}}{e^{P^{1,p}}+e^{P^{2,p}}}) \\ 
- \sum_{q = 1}^{m_2} log_{e} (\frac{e^{P^{2,q}}}{e^{P^{1,q}}+e^{P^{2,q}}}) \\
= - \sum_{p = 1}^{m_1}  log_{e} (e^{P^{1,p}}) - \sum_{q = 1}^{m_2} log_{e} (e^{P^{2,q}}) \\
+ \sum_{p = 1}^{m_1} log_{e}(e^{P^{1,p}}+e^{P^{2,p}})
+ \sum_{q = 1}^{m_2} log (e^{P^{1,q}}+e^{P^{2,q}})
\end{split}    
\end{equation}

where $P^{1,p}$ is the probability of measurement outcome $\sigma_3 = +1 $ on the first qubit for $p$-th sample of the class labelled 1; $P^{2,p}$ is the probability of measurement outcome $\sigma_3 = +1 $ on the second qubit  for $p$-th sample of the class labelled 1; $P^{1,q}$ is the probability of measurement outcome $\sigma_3 = +1 $ on the first qubit for $q$-th sample of the class labelled 2; $P^{2,q}$ is the probability of measurement outcome $\sigma_3 = +1 $ on the second qubit for $q$-th sample of the class labelled 2.
 
\vspace{0.4em}

Minimizing the cross-entropy loss given by equation ~\ref{Derivation1} above involves maximizing $P^{1,p}$ for as many samples (indexed with $p$) as possible belonging to the class labelled 1 and also maximizing $P^{2,q}$ for as many samples (indexed with $q$) as possible belonging to the class labelled 2. Following equation ~\ref{Derivation1}, this minimization can further be enhanced if $P^{2,p}$ is minimized for as as many samples (indexed with $p$) as possible belonging to the class labelled 1 and if $P^{1,q}$ is minimized for as many samples (indexed with $q$) as possible belonging to the class labelled 2. This means that for the samples belonging to the class labelled 1, $\sum_{p = 1}^{m_1}log_{e}(e^{P^{1,p}}+e^{P^{2,p}})$ can remain almost constant over every iteration. Similarly, for all the samples belonging to the class labelled 2, $\sum_{q = 1}^{m_2} log (e^{P^{1,q}}+e^{P^{2,q}})$ can remain almost constant over every iteration. Thus, minimizing only the first two terms in equation ~\ref{Derivation1} leads to the training for the binary classification problem. 

\vspace{0.4em}
Minimizing the first two terms in equation ~\ref{Derivation1} is same as minimizing the following loss function:

\begin{equation}
\label{Derivation2}
{\cal E} = \sum_{p = 1}^{m_1} (1 - P^{1,p}) + \sum_{q = 1}^{m_2} (1 - P^{2,q})
\end{equation}

In order to minimize the loss in equation ~\ref{Derivation2}, for samples (indexed $p$) belonging to the class labelled 1, $P^{1,p}$ should tend towards 1. Similarly, for samples (indexed $q$) belonging to the class labelled 2, $P^{2,q}$ should tend towards 1. But instead of carrying out measurements on the second qubit and obtaining $P^{2,q}$, measurement can be carried out on the first qubit itself for samples (indexed $q$) belonging to the class labelled 2. The $P^{1,q}$ can be minimized instead (tend towards 0). 

Referring to $P^{1,p}$ as $P_+^p$ and to $P^{1,q}$ as $P_+^q$, we obtain the loss function to be:

\begin{equation}
 \label{Derivation3}
     {\cal E} = \sum_{p = 1}^{m_1} (1 - P_+^p) + \sum_{q = 1}^{m_2} (1 - (1 - P_+^q))
 \end{equation}
 , which is same as equation ~\ref{error_lin_bin} in Section III.

\nocite{*}


%

\end{document}